\documentclass[aps,prb,a4paper,reprint,amssymb,amsfonts,superscriptaddress]{revtex4-1}
\pdfoutput=1 

\usepackage{amsmath} 
\usepackage{mathptmx} 
\usepackage{bm}
\usepackage{graphicx}
\usepackage{color}
\usepackage{pstricks} 
\usepackage{verbatim} 
\usepackage{hyperref}
\hypersetup{
pdfborder={0 0 0}, 
colorlinks=true, 
linkcolor=blue,
citecolor=blue,
breaklinks=true, 
urlcolor=blue
}

\newcommand{\gv}[1]{\ensuremath{\mbox{\boldmath$ #1 $}}} 
\newcommand{\grad}[1]{\gv{\nabla} #1} 

\begin{document} 

\title{Helicity within the vortex filament model} 

\author{R. H\"anninen}
\affiliation{Department of Applied Physics, Aalto University, P.O. Box 15100, FI-00076 AALTO, Finland}
\email{Risto.Hanninen@aalto.fi}
\author{N. Hietala}
\affiliation{Department of Applied Physics, Aalto University, P.O. Box 15100, FI-00076 AALTO, Finland}
\author{H. Salman}
\affiliation{School of Mathematics, University of East Anglia, Norwich Research Park, United Kingdom}

\date{June 30, 2016} 



\begin{abstract}
Kinetic helicity is one of the invariants of the Euler equations that is associated with the topology of vortex lines within the fluid. In superfluids, the vorticity is concentrated along vortex filaments. In this setting, helicity would be expected to acquire its simplest form. However, the lack of a core structure for vortex filaments appears to result in a helicity that does not retain its key attribute as a quadratic invariant. By defining a spanwise vector to the vortex through the use of a Seifert framing, we are able to introduce twist and henceforth recover the key properties of helicity. We present several examples for calculating internal twist to illustrate why the centreline helicity alone will lead to ambiguous results if a twist contribution is not introduced. Our choice of the spanwise vector can be expressed in terms of the tangential component of velocity along the filament. Since the tangential velocity does not alter the configuration of the vortex at later times, we are able to recover a similar equation for the internal twist angle for classical vortex tubes. Our results allow us to explain how a quasi-classical limit of helicity emerges from helicity considerations for individual superfluid vortex filaments.
\end{abstract}

\maketitle
%
%


\section{Introduction}\label{s.intro}

Ideal classical fluids possess a number of conserved quantities such as energy, momentum, angular momentum and kinetic helicity. Of these, only energy and kinetic helicity correspond to quadratic invariants in the system. First identified as an invariant for ideal classical fluids by Moffatt \cite{Moffatt1969}, kinetic helicity has been recognized to be related to the topology of the vorticity field in classical fluids. Aside from its relevance to classical fluids, helicity also plays an important role in a number of other contexts such as plasmas, cosmic strings, magnetohydrodynamics, and biology\cite{MoffattPNAS2014,Tkalec2011,Dennis2010,Berry2001,KediaPRL2013,Chichak2004,VachaspatiPRL1994,Bekenstein1992,Sato1995,Kozhevnikov1995,Faddeev1997}. 

If vorticity is concentrated within vortex tubes, the helicity is in part determined by the self-knotting and linking of the vortices
within a tangle. In recent years, it has become possible to realize particular cases of knotted vortex tubes produced in water and to study them under controlled conditions in the laboratory\cite{Kleckner2013}. This has paved the way to test assertions arising from considerations involving helicity. Recently, tying a simple knot has been experimentally demonstrated even in a Bose-Einstein condensate of an ultracold atomic Bose gas\cite{Hall2016}. The characteristics of superfluid vortices in such Bose gases are essentially identical to vortices that we would expect to form in superfluid $^4$He. Therefore, similar knotted vortex excitations would be expected to arise in superfluid helium although their experimental verification would be more of a challenge since it is only recently that direct visualization of quantized vortices in $^4$He has become possible using small tracer particles \cite{BewleyNature2006}.

In contrast to classical fluids, in superfluids, the vorticity is concentrated along line filaments and the circulation is quantized in units of Planck's constant. Moreover, superfluid vortices are not subject to diffusive processes that are present in classical fluids.
Yet, although superfluid vortices appear to resemble vortices of ideal classical fluids, their topology is not frozen into the flow because quantum effects can act to allow vortices to undergo reconnections. In other words, superfluid vortices are not constrained by the Helmholtz laws of vortex motion. Given these observations, it appears that helicity would acquire its simplest form in a superfluid. At the same time, reconnections delineate the boundaries of sharply defined intervals involving constant vortex topologies. This makes superfluids an ideal setting in which to understand properties of helicity and how it relates to the topology of the superfluid vortex filaments. 

Numerical studies of knotted vortex structures have previously been performed using both a hydrodynamic description based on a vortex filament model\cite{MaggioniPRE2010} and a microscopic mean field model in the form of the Gross-Pitaevskii equation that can directly resolve the core structure\cite{Proment}. In superfluids, the core of a vortex is characterised by a depletion of the density field which vanishes along the centreline of the vortex\cite{note3He}. Therefore, although superfluid vortices have a finite core size characterized by the rapid variation of the density field, the vorticity remains singular and concentrated along a filament due to the divergent nature of the velocity within the core. When the core size is much smaller than any other characteristic scale and compressiblity effects can be neglected, one may adopt a vortex filament model in which the dynamics of the superfluid reduces to studying how a vortex or a tangle of vortices evolves. Both these conditions are typically satisfied in $^4$He. In this case, the vortex core size is of the order of 1 {\AA}. Moreover, since superfluid helium is a liquid, it is much more reasonable to assume that it behaves as an incompressible fluid. The vortex filament model has served as an excellent description to study the dynamics of vortices in this systems.

Since a vortex filament has no internal structure associated with the vorticity field, there is no notion of the twisting of the vortex lines within the core of a vortex as is the case for vortex tubes with a cross-sectional area characterising finite extent of the vorticity field. Such a scenario leads to a number of complications when evaluating helicity for a vortex filament. In particular, it would appear that helicity conservation can not be satisfied, even in between reconnection events. 

In this work, we will show that without introducing the notion of an internal twist, helicity conservation can not be satisfied within the vortex filament model. To introduce internal twist, a spanwise vector must be defined along the length of the filament. We will show that a natural choice is to set the spanwise vector to coincide with a direction of constant velocity potential upon which we are able to recover the key properties of helicity. In superfluids the quantum mechanical phase of the order parameter plays the role of the velocity potential. We note that similar choices have been suggested also other authors\cite{Baggaley2014,Kleckner2016,ZuccherRicca2015,Brachet2016}. We will illustrate that this choice of the spanwise vector coincides with working in the so-called Seifert frame in which case helicity becomes trivially zero \cite{Seifert1985,Akhmetev1992}. In this frame, the direction of the spanwise vector is related to the tangential component of standard Biot-Savart velocity along the vortex filament. When the filament is interpreted as a limiting case of a classical vortex tube endowed with internal structure, the tangential velocity of the vortex filament is modified. Since the tangential velocity does not affect the overall configuration of the vortices, our definition remains consistent with the vortex dynamics produced by a vortex filament model. This allows us to generalize our results for superfluid vortices to classical vortices. In particular, our results will allow us to establish how a quasiclassical limit of helicity can emerge from the microscopic quantum description.

\section{Results}

\subsection{Helicity and its components}\label{s.helicity}

We will begin by recalling some key results concerning helicity in classical fluids.
Helicity in fluid dynamics is defined as
\begin{equation}
\mathcal{H} = \int {\bf v}\cdot {\bm \omega} \; d^3 {\bf r} \, , \label{e.helicity}
\end{equation}
where the integration is over the whole space where the flow is defined, ${\bf v}={\bf v}({\bf r},t)$ is the velocity field, 
and ${\bm \omega}({\bf r},t) = \grad \times {\bf v}$ is the 
corresponding vorticity field. 
If the vorticity is taken to be concentrated within vortex tubes but is zero otherwise, 
the helicity will have a topological interpretation. For simplicity, we will consider the scenario 
where each vortex tube carries the same vorticity flux $\kappa$. 
This flux can be identified with the circulation which for quantized vortices is given by $\kappa = h/m$. 
Here, $h$ is Planck's constant and $m$ is the atomic mass (e.g.\ of the $^4$He atom).
Following Moffatt and Ricca\cite{MoffattRicca1992}, we decompose the helicity into the linking number $L = \sum_{i\neq{j}}L_{ij}$ 
(where the double summation is over different vortices and $L_{ij}$ is the Gauss linking number), the writhe $W=\sum_{i}W_i$, 
and the twist part of the helicity $T = \sum_{i}T_{i}$ such that
\begin{equation}
\mathcal{H} = \kappa^2(L + W + T) \, . \label{e.Hparts}
\end{equation}

To provide more explicit expressions for these terms, we parameterize a vortex in terms of its arclength $\xi$ so that the position vector of a vortex filament is given by ${\bf s} = {\bf s}(\xi,t)$. It is then natural to introduce the Frenet-Serret basis given by Eqs.\ \eqref{eqn_SF} to evaluate the curvature $c(\xi,t)$, and torsion $\tau(\xi,t)$ of each vortex filament:
\begin{eqnarray}
{\bf s}' = \hat{{\bf t}}, \;\;\;\;\;  
\hat{{\bf t}}' = c \hat{{\bf n}}, \;\;\;\;\; 
\hat{{\bf n}}' = -c \hat{{\bf t}} + \tau \hat{{\bf b}}, \;\;\;\;\; 
\hat{{\bf b}}' = -\tau \hat{{\bf n}}, \label{eqn_SF}
\end{eqnarray}
where $\hat{\bf n}$ and $\hat{\bf b}$ are local normal and bi-normal (unit vectors), respectively. Here, primes imply differentiation with respect to arclength. Denoting the centreline of the vortex tube $i$ with position vector ${\bf s}$ by $C_i$, we can then identify a unit span-wise vector ${\bf N}$ that determines the twisting of the vortex lines within the vortex tube such that
\begin{equation}\label{e.Neq}
{\bf N} = \hat{\bf n}\cos\theta+\hat{\bf b}\sin\theta.
\end{equation}
This vector can be attributed to a ribbon like structure that is oriented with respect to the unit normal and and bi-normal vectors. 
With these definitions, the different helicity components can be written as\cite{MoffattRicca1992,Fuller1971}
\begin{eqnarray}
L &=& \frac{1}{4\pi} \sum_{i\neq{j}} \oint_{C_i} \oint_{C_j}\frac{({\bf s}_1-{\bf s}_2) \cdot d{\bf s}_1\times d{\bf s}_2}{|{\bf s}_1-{\bf s}_2|^3} \, , \label{e.linknum} \\ 
W &=& \frac{1}{4\pi} \sum_i \oint_{C_i} \oint_{C_i}\frac{({\bf s}_1-{\bf s}_2) \cdot d{\bf s}_1\times d{\bf s}_2}{|{\bf s}_1-{\bf s}_2|^3} \, , \label{e.writhe}  \\ 
T &=& \frac{1}{2\pi}\sum_i \oint_{C_i}({\bf N}\times{\bf N}')\cdot d{\bf s} \, . \label{e.twist}
\end{eqnarray}
The use of the Frenet-Serret equations results in a twist contribution to the helicity that is given by
\begin{equation}\label{e.Twist}
T = \frac{1}{2\pi}\sum_i\oint_{C_i}\left(\tau+\frac{d\theta}{d\xi}\right) d\xi
  = T_{\rm tors}+\frac{1}{2\pi}\sum_i\lbrack\theta\rbrack_{C_i} \, . 
\end{equation}
Here the torsion part of the twist $T_{\rm tors}$ can be evaluated using the local torsion $\tau(\xi)$. 
The last term in Eq.\ \eqref{e.Twist} describes the internal (or intrinsic) twist $T_{\rm tw}$. 
We note that, since ${\bf N}$ is defined in terms 
of the normal and bi-normal unit-vectors, a constant value of ${\bf N}$ only implies that the 
total twist is zero. The internal twist can still be nonzero if torsion is also nonzero.

For closed vortex tubes, both the Gauss linking number and the internal twist can only take integer values.
The writhe, the Gauss linking number, and the torsion part of the twist do not depend on the way one prescribes 
the spanwise vector ${\bf N}$. However, the internal twist depends on the choice for ${\bf N}$ \cite{BanchoffWhite1975}. 
Nevertheless, once ${\bf N}$ is prescribed, the self-linking number for vortex $i$, which equals $W_i+T_i$, also becomes 
a topological invariant. It follows that by redefining ${\bf N}$ we may designate different values to the helicity.

Theoretically it has been shown that under continuous deformation of the vortex tube, writhe $W$ and 
total twist $T = T_{\rm tors}+T_{\rm tw}$ vary continuously. However, if at some time $t=t_c$ a curve 
passes through an inflectional configuration, i.e., its curvature $c(\xi)$ vanishes at some point $\xi=\xi_c$, 
then the torsion part of the helicity $T_{\rm tors}$ develops a jump of $\pm 1$. At the instant when the curve has
a point with zero curvature, the local torsion, $\tau$, diverges around $\xi=\xi_c$, but this divergence is integrable. 
The resulting jump is compensated by a jump in the internal twist of $\mp 1$.

\subsection{Helicity for vortex filaments}\label{s.hvfm}

When the superfluid is described by a complex wavefunction (order parameter $\Psi$) then
vorticity can arise only in the form of line-like topological defects. For pure line vortices 
(i.e.\ neglecting other excitations in the superfluid), the corresponding velocity
potential $\varphi$ can be identified with the phase of the 
order parameter such that $\Psi=\sqrt{\rho/m}\exp({\rm i}\varphi)$. In this case, the phase is well defined everywhere, except 
at the vortex cores where the vorticity is non-zero. 

If we ignore compressibility effects and focus on the incompressible motion given by the vortices that correspond to 
pure line defects, the superfluid velocity can be calculated using the standard Biot-Savart law 
\begin{equation}\label{e.BS}
{\bf v}_{\rm BS}({\bf r},t) = \frac{\hbar}{m}\grad \varphi = \frac{\kappa}{4\pi}\oint\frac{({\bf s}_1-{\bf r})\times d{\bf s}_1}
{\vert {\bf s}_1-{\bf r}\vert^3}\, , 
\end{equation}
where the integration is carried out along all the vortex lines. 
This equation can be directly applied at any point ${\bf r}\neq {\bf s}$ (i.e.\ not lying on a vortex). 
At a point ${\bf r} = {\bf s}$, the divergent integrand implies that the normal and binormal components of the velocity are not well defined at the vortex line. In contrast, the tangential component of ${\bf v}_{\rm BS}$ remains well-behaved. This difficulty can be circumvented by regularizing the integral in the vicinity of the point on a vortex as will be described in Sec.\ \ref{s.dvfm}.

If the Biot-Savart integral is inserted into the Eq.\ \eqref{e.helicity} for the helicity, we appear to recover only the linking number and the writhe terms. Therefore, by using the above expression for the
Biot-Savart velocity it seems that the helicity is not the same as the one presented earlier in 
Sec.~\ref{s.helicity}. 

Although a vortex filament has no internal core structure, on physical grounds a vortex is endowed with a scalar field, the velocity potential that permits us to identify a spanwise vector to a direction of constant phase\cite{NoteIncomp}. The direction of the constant phase (or velocity potential) of the complex wavefunction has also been proposed by others as a possible direction that can be used to define ${\bf N}$ for superfluid vortices\cite{Baggaley2014,Kleckner2016,ZuccherRicca2015}. 
This choice for the ${\bf N}$ turns out to be equivalent to using the so-called Seifert framing\cite{NoteSeifertFrame}. 
This particular framing is special because the total helicity 
is always zero when evaluated in this frame\cite{Seifert1985,Akhmetev1992}. 
With ${\bf N}$ as defined, it is now possible to recover a torsion and internal twist contribution to helicity. 
To see how this arises, we note that in the Seifert framing we have
\begin{eqnarray}
\lim_{\epsilon\rightarrow{0}} {\bf N}\cdot\nabla{\varphi}\vert_{{\bf r}={\bf r}^*(\xi^*)} = 0\, ,
\end{eqnarray}
where ${\bf r}^* = {\bf s}+\epsilon {\bf N}$ represents a point located near a point ${\bf s}$ lying on the vortex.
Here $\nabla\varphi$ is proportional to the superfluid velocity, which is normal to surfaces of 
constant phase near a vortex. By construction, 
on the curve $C^*$ given by ${\bf r}={\bf r}^*$ the phase is constant so that we can write 
$\varphi({\bf r}^*(\xi^*)) = \varphi_0$, where $\varphi_0$ is an arbitrary constant. 
This implies that
\begin{eqnarray}
\!\!\!\!\!\!\frac{d\varphi\vert_{{\bf r}\!=\!{\bf r}^*(\xi^*)}}{d\xi^*} = 
\nabla\varphi\vert_{{\bf r}={\bf r}^*(\xi^*)}\cdot \hat{\bf t}^*\!=0, \;\; \Rightarrow  
{\bf v}_{\rm BS}({\bf r}^*)\cdot \hat{\bf t}^*\!= 0\, ,
\label{e.vtstar}
\end{eqnarray}
where $\hat{\bf t}^* = d{\bf r}^*/d\xi^*$ is the tangent for the curve $C^*$. In the limit 
$\epsilon\rightarrow{0}$, $C^*$ and $C$ coincide and the tangent vector $\hat{{\bf t}}^*$ then 
approaches the tangent $\hat{\bf t}$ of the vortex line $C$. 
However, the above cannot be used to deduce that the tangential Biot-Savart velocity along the vortex is zero. This is because, 
even if the azimuthal component (component around the vortex centreline) of $\hat{\bf t}^*$ goes to zero, the azimuthal 
component of the velocity diverges. Therefore, a more detailed analysis is required to determine the tangential velocity 
along the filament. By using the chain rule we can express $\hat{\bf t}^*$ as
\begin{eqnarray}
\hat{\bf t}^* = \frac{d{\bf r}^*}{d\xi^*}
 = \left(\frac{d{\bf s}}{d\xi}+\epsilon\frac{d{\bf N}}{d\xi}\right)\frac{d\xi}{d\xi^*} 
 = \left(\hat{\bf t}+\epsilon\frac{d{\bf N}}{d\xi}\right)\frac{d\xi}{d\xi^*} \, .\,\,\,\,
\end{eqnarray}
After inserting the definition for ${\bf N}$ given by Eq.~\eqref{e.Neq}, and applying the Frenet-Serret 
equations, plus noting that the local azimuthal direction around the vortex tangent is given by 
$\hat{\bf e}_\theta = -\sin\theta\hat{\bf n}+\cos\theta\hat{\bf b}$,
the above equation for $\hat{\bf t}^*$ simplifies to
\begin{eqnarray}
\hat{\bf t}^*
 &=& \left(\left(1-\epsilon{c}\cos\theta\right)\hat{\bf t}
+\epsilon \left(\tau+\frac{d\theta}{d\xi}\right)\hat{\bf e}_\theta\right)\frac{d\xi}{d\xi^*} 
\nonumber \\
 &=& \frac{ \left(1-\epsilon{c}\cos\theta\right)\hat{\bf t}
 +\epsilon k \hat{\bf e}_\theta }{ \sqrt{(1-\epsilon{c}\cos\theta)^2+\epsilon^2k^2} }
\label{e.tstar}
\end{eqnarray}
where $k = k(\xi)= \tau+\frac{d\theta}{d\xi} = ({\bf N}\times{\bf N}')\cdot{\hat{\bf t}}$ 
describes the rate of rotation of the spanwise vector around the local tangent $\hat{\bf t}$. 
In arriving at the final expression given in Eq.~\eqref{e.tstar}, we have assumed that the local curvature satisfies the condition $c \ll 1/\epsilon$, and we have used the unit normalization to determine the value for the common factor, $d\xi/d\xi^*$.

We would expect that the leading order terms for the Biot-Savart velocity at ${\bf r} = {\bf r}^*$ are given by
\begin{eqnarray}
{\bf v}_{\rm BS}({\bf r}^*) = \frac{\kappa}{2\pi\epsilon}\hat{\bf e}_\theta+({\bf v}_{\rm BS}\cdot\hat{\bf t})\hat{\bf t} \, .
\label{e.vsrstar}
\end{eqnarray}
Now substituting Eqs.~\eqref{e.tstar} and \eqref{e.vsrstar} into 
Eq.~\eqref{e.vtstar} and taking the limit $\epsilon\rightarrow{0}$, we obtain that
\begin{eqnarray}
\label{e.itaSeifert}
{\bf v}_{\rm BS}\cdot\hat{\bf t} = -\frac{\kappa}{2\pi}({\bf N}\times{\bf N}')\cdot{\hat{\bf t}} 
= -\frac{\kappa}{2\pi}\left(\tau(\xi)+\frac{d\theta}{d\xi}\right)
\, . 
\end{eqnarray}
This implies that we may obtain the internal twist angle $\theta$ after calculating the
tangential component of the Biot-Savart velocity and the local torsion. 

To clarify the relation between classical and quantum vortices, in order to establish how a quasiclassical limit can arise from the quantum case, let us consider the scenario of a classical vortex tube with a finite but small cross-sectional area. As shown by Moffatt and Ricca\cite{MoffattRicca1992}, for a vortex with a finite core, it is possible to relate the twist of the vorticity flux lines to the twist of the spanwise vector in the equation for the helicity. Thus, in contrast to a quantum vortex, for a classical vortex, the twist $T$, defined in Eq.~\eqref{e.Twist}, originates from the azimuthal component of the vorticity. 

In the case when the characteristic length scale along the vortex tube is much larger than the vortex core, it may be convenient to use a filament representation and study the dynamics of how the centreline vorticity evolves. However, this filament description of a vortex tube follows from a scale separation assumption rather than being an intrinsic representation of the actual vorticity field. A key difference with a superfluid vortex is that the internal vorticity structure now gives rise to a component of velocity along the centreline of the vortex tube.
The total velocity for a vortex filament used to model a thin vortex tube is therefore given by 
\begin{equation}\label{e.vtang}
{\bf v}({\bf s}) = {\bf v}_{\rm BS}({\bf s}) + v_{\rm twist}({\bf s})\hat{\bf s}' \, ,
\end{equation}
The additional tangential twist velocity arising from the twisting of the vortex flux lines is given by 
\begin{equation}\label{e.vtwist}
v_{\rm twist}({\bf s}) = \frac{\kappa}{2\pi}({\bf N}\!\times\!{\bf N'})\cdot \hat{\bf s}' 
           = \frac{\kappa}{2\pi}\left(\tau(\xi)+\frac{d\theta}{d\xi}\right) \, .
\end{equation}
Here $\tau(\xi)$ is the local torsion at $\xi$ and ${\bf N}$ describes the rotation of the 
vorticity fluxlines inside the vortex core, as illustrated in Fig.~\ref{f.twistKW}. 
If the cross-sectional area of the vortex tube is assumed to remain constant along its length, then together with 
the assumption of incompressibility the $v_{\rm twist}$ velocity would need to be constant 
on each vortex\cite{noteContCore}. The helicity conservation would then fix the time dependence of $v_{\rm twist}$. 

We can now see that if the total tangential velocity is zero everywhere on the 
vortex, we recover Eq.~\eqref{e.itaSeifert} for the internal twist angle that was obtained by using the Seifert frame for superfluid vortex filaments. Since the vorticity of a superfluid vortex does not have an azimuthal component, superfluid vortices do not have an additional intrinsic degree of freedom that can be ascribed to classical vortices. Therefore, in the Seifert frame, the helicity of a superfluid vortex is always zero. In contrast, ${\bf N}$ is determined by the structure of the vorticity field within the core of a classical vortex and, as such, the helicity is not necessarily the same as for a superfluid vortex. 

By assuming that the helicity would remain zero also for the classical case, and that the twist velocity remains constant along the filament as required by incompressibility, the internal twist angle for vortex $i$ is obtained using the following equation:
\begin{equation}
\frac{d\theta_i}{d\xi} \!=\! -\tau(\xi)-\!\frac{2\pi}{\kappa{\mathcal{L}_i}}\int_{C_i}\!{\bf v}_{\rm BS}\cdot d{\bf s} 
=\!-\tau(\xi)-\!\frac{2\pi}{{\mathcal{L}_i}}\left(\!\sum_{j\neq{i}}L_{ij}+W_i\!\right) \, .
\label{e.itaVFM}
\end{equation} 
Here $\mathcal{L}_i=\int_{C_i}d\xi$ denotes the length of vortex $i$. 
If we want that the helicity is simply given by linking, one should only replace ${\bf v}_{\rm BS}$ 
in Eq.~\eqref{e.itaVFM} with the velocity caused by vortex $i$ alone, i.e., set $L_{ij} = 0$.

Both Eqs.~\eqref{e.itaSeifert} and \eqref{e.itaVFM} give zero helicity and the same value for 
the total twist $T$. The difference appears only in the local value of the internal twist angle,
i.e., the frame in the classical case is generally different from the Seifert case since it is associated with the structure of the vorticity within the core.

\subsection{Straight vortex with a Kelvin wave}\label{s.vortexKW}

\begin{figure}[!t]
\centerline{\includegraphics[width=0.85\linewidth]{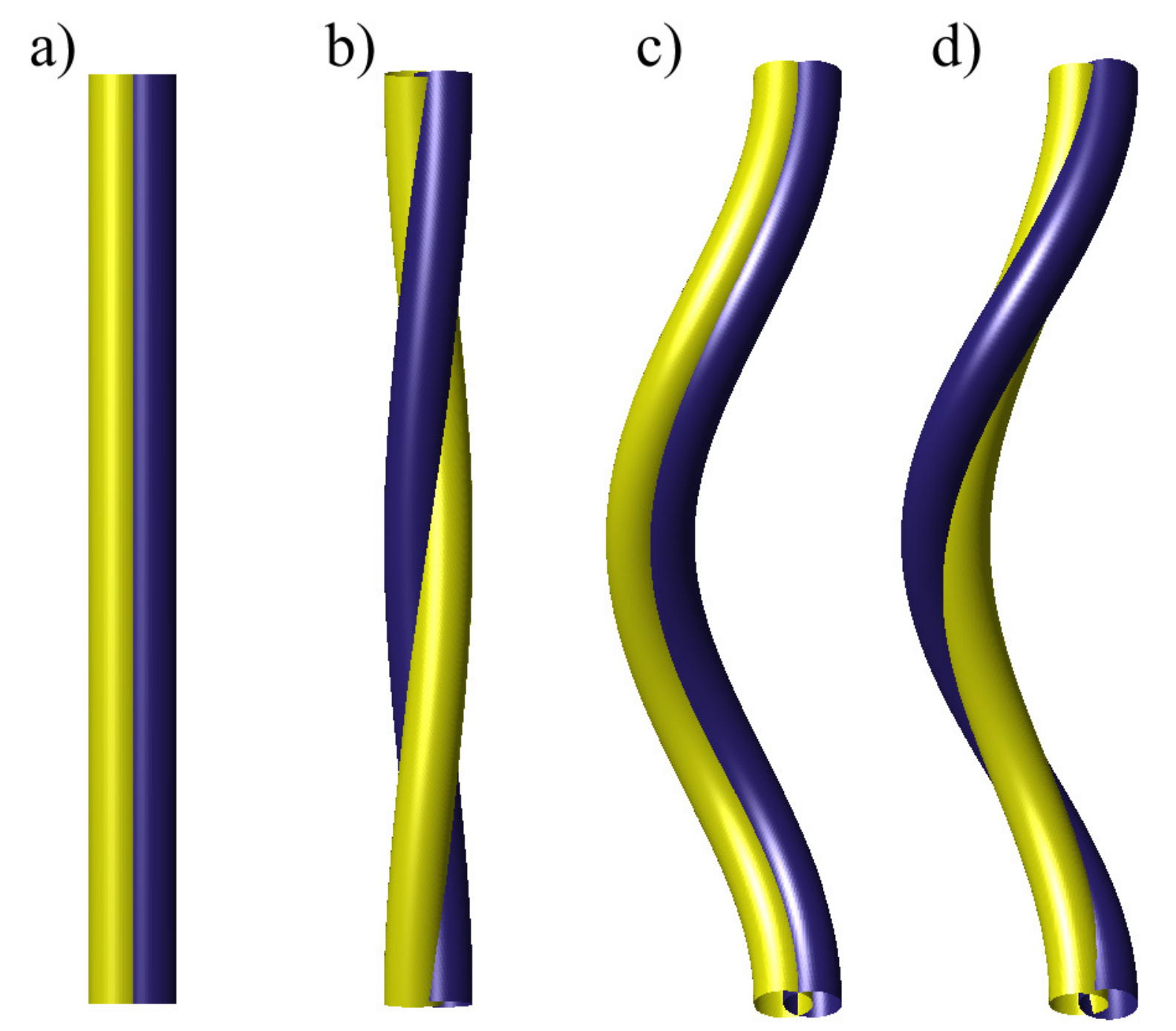}}
\caption{
Illustration of the effect of the twist on nearby vortex tubes. The twist of the two tubes describes the rotation of the unit spanwise vector ${\bf N}$ around the vortex centreline. This direction would coincide with the rotation of the vorticity flux lines inside the vortex core of a classical vortex. For a superfluid, this vector coincides with the direction of the constant phase of the order parameter deep within the vortex. For a straight vortex, the contribution to the helicity in a) is zero while in b) the nonzero (internal) twist contributes to helicity. For a Kelvin wave in c) the internal twist (nonzero) is set to cancel the writhe and the torsion part of the twist, which in d) gives a nonzero helicity. In the limit of zero Kelvin-wave amplitude the configuration d), where the internal twist is zero, results a configuration b), while the configuration c) reduces to a).}
\label{f.twistKW}
\end{figure}

Initially, we will focus on idealised vortex configurations to elucidate the different properties of helicity in order to facilitate our understanding of how helicity behaves in more realistic situations. A straight vortex with only an azimuthal velocity field around its core has zero helicity.
We will, therefore, consider perturbations on an otherwise straight vortex in
the form of a helical Kelvin wave with amplitude $A$, i.e.: 
\begin{equation}
x = A \cos(kz)\,, \hspace{4mm} y = A \sin(kz)\,, \hspace{4mm} z = \frac{\xi}{\sqrt{1+A^2k^2}} \,.
\end{equation}      
The last equation relates $z$ to the arc length $\xi$. The unit tangent $\hat{\bf t}=\hat{\bf s}'$, 
principal normal $\hat{\bf n}$, and bi-normal $\hat{\bf b}$ are then given by
\begin{eqnarray}
\hat{\bf t} &=& -\frac{Ak\sin(kz)}{\sqrt{1+A^2k^2}}\hat{\bf x} 
               +\frac{Ak\cos(kz)}{\sqrt{1+A^2k^2}}\hat{\bf y}
               +\frac{1}{\sqrt{1+A^2k^2}}\hat{\bf z} \nonumber \\
\hat{\bf n} &=& -\cos(kz)\hat{\bf x}-\sin(kz)\hat{\bf y} \\
\hat{\bf b} &=& \frac{\sin(kz)}{\sqrt{1+A^2k^2}}\hat{\bf x} 
               -\frac{\cos(kz)}{\sqrt{1+A^2k^2}}\hat{\bf y}
               +\frac{Ak}{\sqrt{1+A^2k^2}}\hat{\bf z} \,.\nonumber
\end{eqnarray}
In this case, the local torsion is constant and equal to
$\tau = -\hat{\bf n}\cdot \frac{d\hat{\bf b}}{d\xi} = k/(1+A^2k^2)$, as would be expected for a 
simple helix. The value for the writhe (which can be computed as described in [\onlinecite{Scheeler2014}]) 
and the torsion (per unit period of $\lambda=2\pi/k$) are
\begin{equation}\label{e.TtorsKW}
W = 1-\frac{1}{\sqrt{1+A^2k^2}}\, , \hspace{4mm} T_{\rm tors} = \frac{1}{\sqrt{1+A^2k^2}} \, .
\end{equation}
The value for the writhe implies that the tangential component of the Biot-Savart velocity is constant and given by
\begin{equation}\label{e.vBStangKW}
{\bf v}_{\rm BS}\cdot\hat{\bf s}' = \frac{\kappa{k}}{2\pi}\frac{\sqrt{1+A^2k^2}-1}{1+A^2k^2} \, .
\end{equation}
Using Eq.~\eqref{e.itaSeifert}, the direction of the constant phase is given by 
\begin{equation}
\theta = \theta_0-\frac{k\xi}{\sqrt{1+A^2k^2}} = \theta_0-kz \, .
\end{equation}
Consequently, for helicity to be zero, the internal part of the twist becomes $T_{\rm tw}=-1$ (per period). 
We note that the formulation given by Eq.~\eqref{e.itaVFM}, obtained by using the (constant) twist 
velocity, gives exactly the same value for the internal twist angle as the formulation using the Seifert frame, Eq.~\eqref{e.itaSeifert}.
By setting the integration constant $\theta_0 = \pi$, we can show that the spanwise vector becomes
\begin{eqnarray}
{\bf N} &=& \left(\cos^2(kz)+\frac{\sin^2(kz)}{\sqrt{1+A^2k^2}}\right) \hat{\bf x} \\
        &+&\left(1-\frac{1}{\sqrt{1+A^2k^2}}\right)\cos(kz)\sin(kz) \hat{\bf y} 
        +\frac{Ak\sin(kz)}{\sqrt{1+A^2k^2}}\hat{\bf z} \nonumber \, .
\end{eqnarray}

Therefore, for small amplitudes ${\bf N}\approx\hat{\bf x}$. Moreover, we see that ${\bf N}$ does not rotate around the $z$-axis, 
as shown in panes a) and c) of Fig.~\ref{f.twistKW}, where the helicity is zero. The internal twist originates from the rotation 
of the normal and bi-normal vectors relative to the spanwise vector.

\begin{figure}[!t]
\centerline{\includegraphics[width=0.98\linewidth]{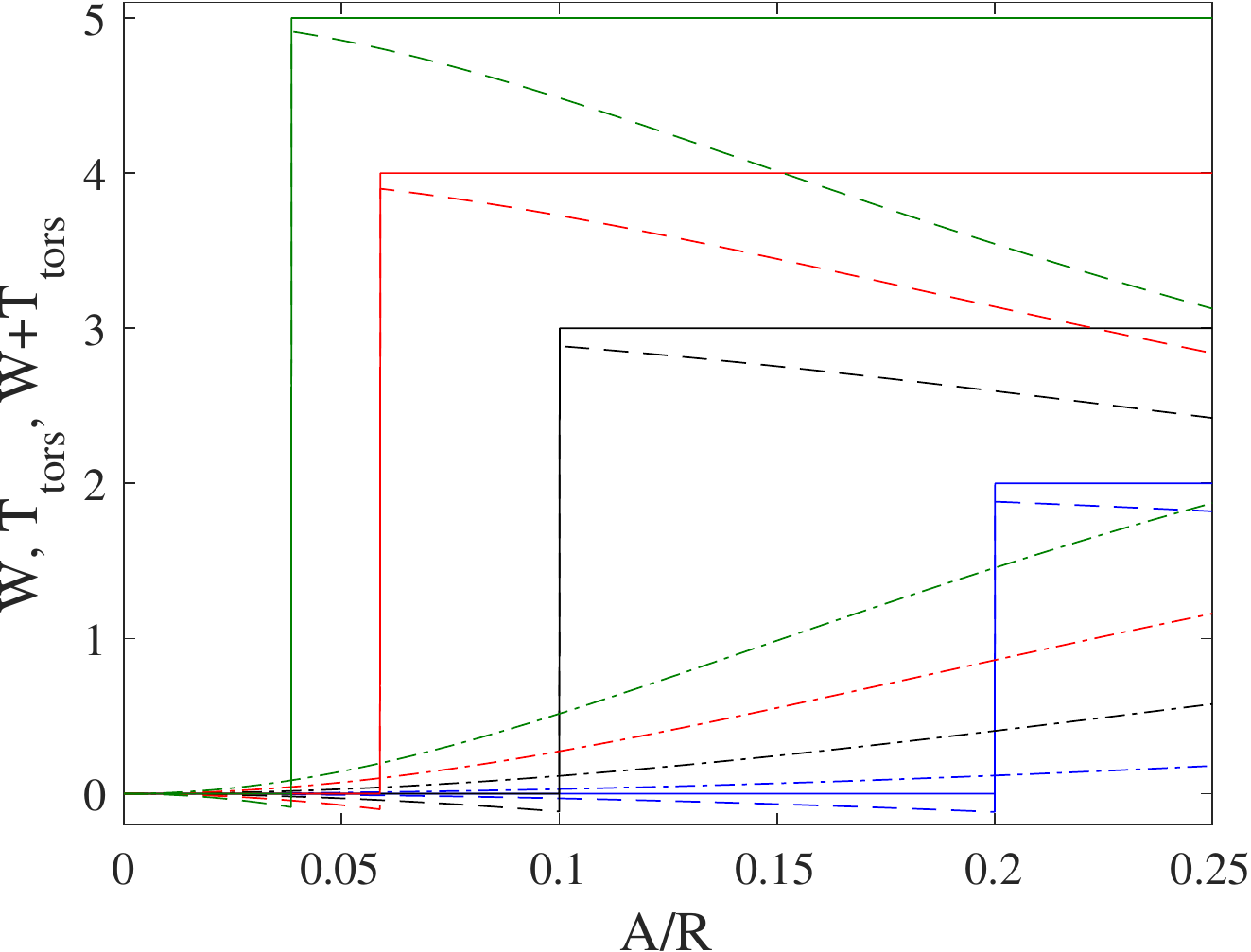}} 
\caption{
Helicity components for a vortex ring of radius $R$ which is occupied by a Kelvin wave with an amplitude of $A$. The different line types describe the writhe, $W$, (dash-dotted) and torsion part of the twist, $T_{\rm tors}$, (dashed) plus their sum (solid). The different family of curves correspond to the Kelvin mode of $m$ = 2 (blue), 3 (black), 4 (red), and 5 (green). The inflection points occur at $A_{\rm c}/R = 1/(m^2+1)$ where $T_{\rm tors}$ jumps by $m$. This jump is compensated by the internal twist $T_{\rm tw}$, which is zero for $A<A_{\rm c}$ and $-m$ for $A>A_{\rm c}$, thus ensuring that the total helicity remains equal to zero for all amplitudes. 
}
\label{f.writhetorsKWring}
\end{figure}

\subsection{Vortex ring with a Kelvin wave}\label{s.ringKW}

Next, we consider a superfluid vortex ring with one Kelvin mode, $m$, where the configuration is given by
\begin{eqnarray}
x &=& (R+A\cos(m\phi))\cos(\phi) \nonumber \\
y &=& (R+A\cos(m\phi))\sin(\phi) \label{e.ringKW} \\
z &=& -A\sin(m\phi) \, . \nonumber 
\end{eqnarray}
Here $\phi$ is the azimuthal angle of a cylindrical coordinate system, $R$ is the ring radius, and $A$ is the 
amplitude of the Kelvin wave. As the amplitude of the Kelvin wave is increased from zero, the standard Biot-Savart 
velocity ${\bf v}_{\rm BS}$ produces a nonzero component along the vortex tangent. On the other hand, if there are no other vortices threading the vortex ring, the helicity should remain zero for any Kelvin wave amplitude.

\begin{figure}[!t]
\centerline{\includegraphics[width=0.99\linewidth]{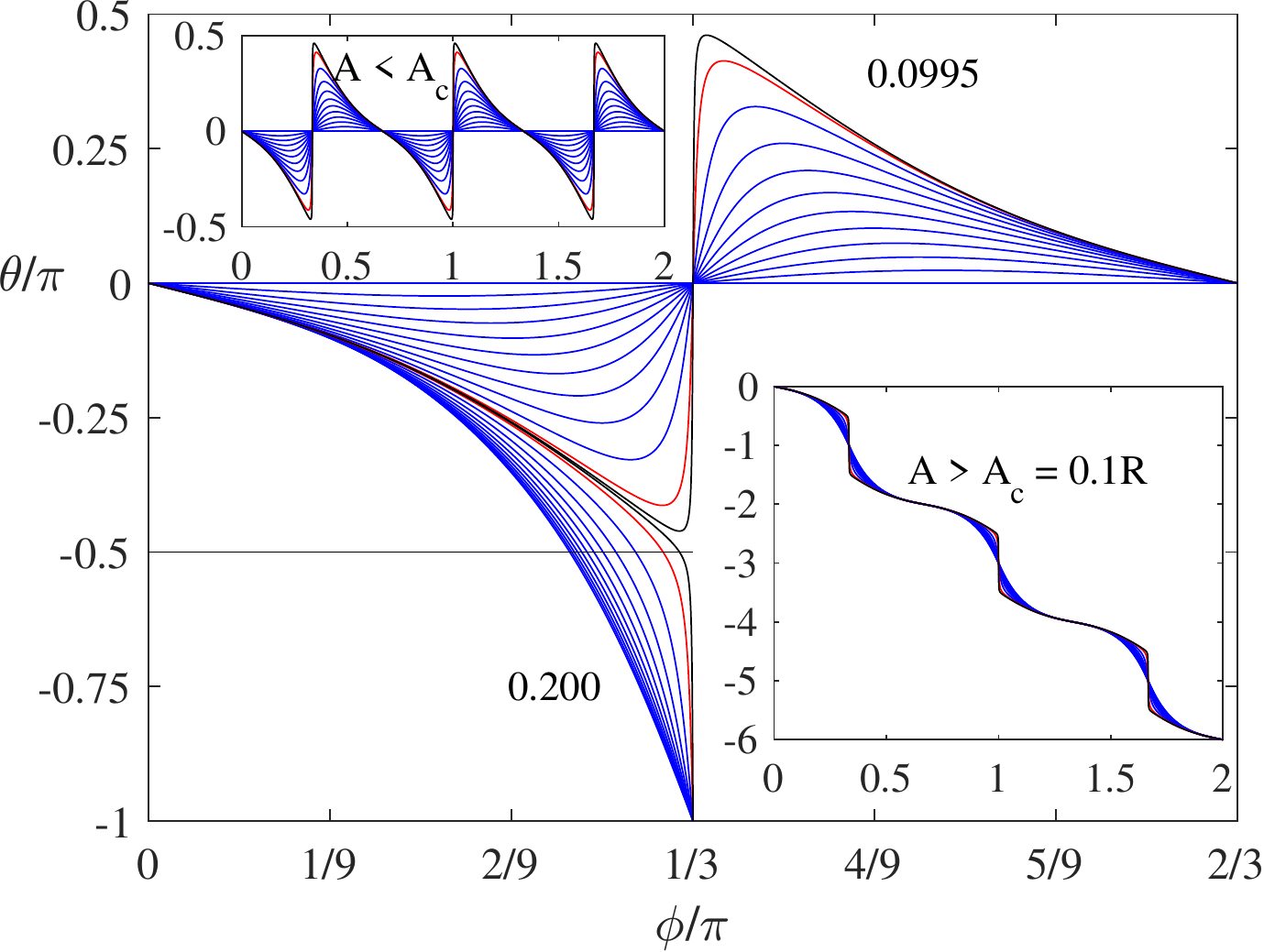}}
\caption{
Internal twist angle for a vortex ring with a Kelvin mode $m$ = 3. The twist angle $\theta$ is plotted as a function of the azimuthal angle $\phi$ when the spanwise vector is chosen to lie along the direction of constant phase. In the {\it top inset} the Kelvin amplitude $A$ is below the critical value of $A_{\rm c}/R = 0.1$, and takes the values $A/R = 0, 0.01, 0.02,\ldots,0.09$ (blue ones), 0.0975 (red) and 0.0995 (black). In the {\it bottom inset} the amplitude is above the critical one and the curves correspond to amplitudes 0.1005 (black), 0.1025 (red) and $0.11, 0.12,\ldots,0.20$ (blue ones) and the total internal twist corresponds to $T_{\rm tw} = -3$. The {\it main figure} shows a zoomed view illustrating the rapid change around $\phi=\pi/3$ that occurs due to the appearance of the inflection points when the Kelvin amplitude approaches the critical value $A_c$. 
}
\label{f.intwistKWring}
\end{figure}

However, if we evaluate the local torsion and the writhe we notice that at small amplitudes they exactly cancel each other. 
This implies that the internal twist $T_{\rm tw}$ as defined must be equal to zero in order for the helicity to remain zero. 
However, at a critical amplitude\cite{noteAc} of $A_c = R/(m^2+1)$, $m$ separate inflection points appear where the curvature 
vanishes and the torsion part of the twist jumps by $m$ due to divergences in the local torsion. This behaviour is illustrated 
in Fig.~\ref{f.writhetorsKWring}, where we have plotted the writhe and torsion part of the helicity, plus their sum as a 
function of the Kelvin wave amplitude. This key observation shows that, for $A>A_c$, the internal twist must give rise to a 
compensating contribution of $T_{\rm tw} = -m$ to conserve helicity. These observations are also consistent with the predictions 
of Moffat and Ricca\cite{MoffattRicca1992} obtained for a classical vortex tube.

\begin{figure}[!t]
\centerline{\includegraphics[width=0.93\linewidth]{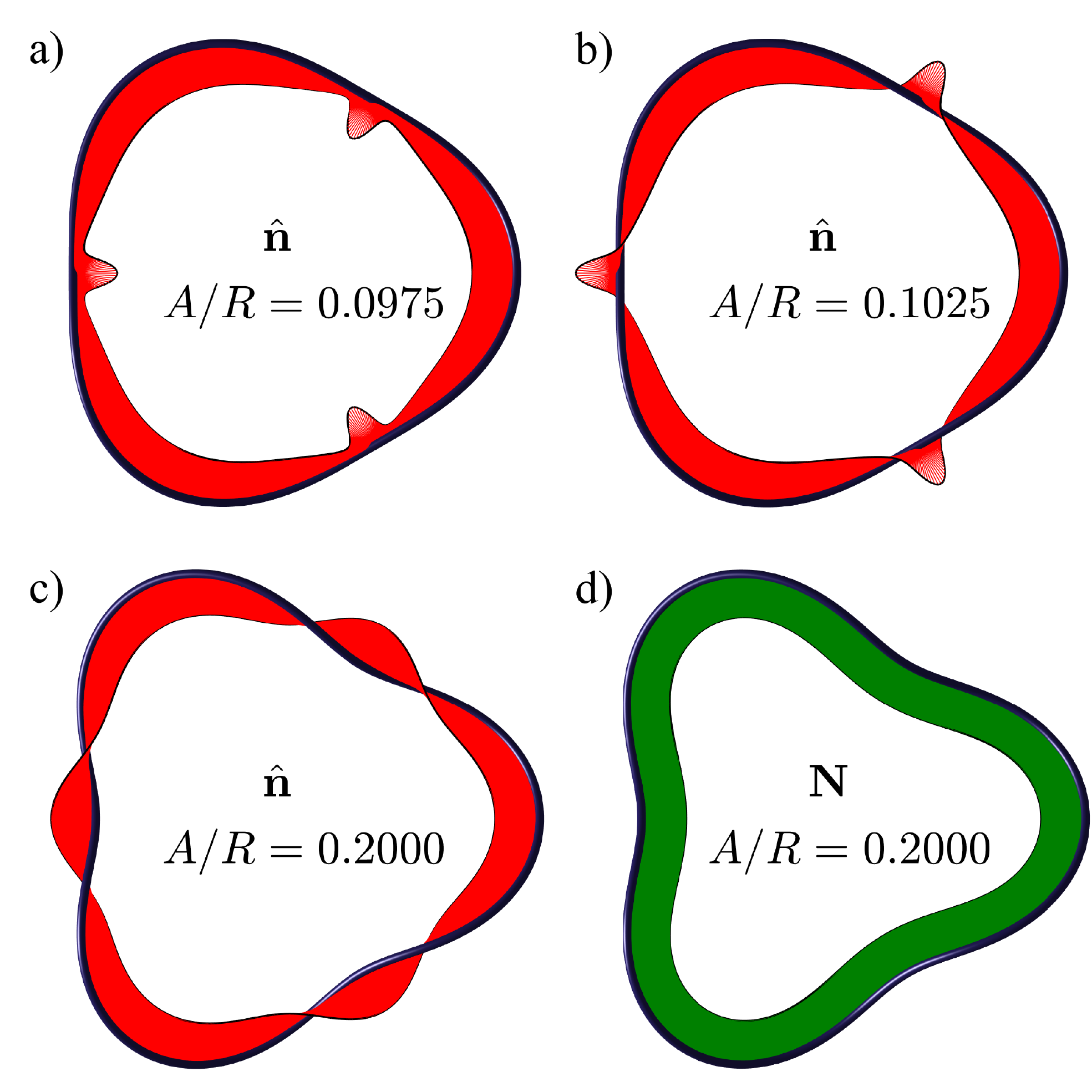}}
\caption{
Vortex ring with a Kelvin mode of $m = 3$. The blue tubes denote the vortex and the red strips on panels a), b), and c) denote the normal unit vectors where the ratio of the Kelvin wave amplitude to the vortex ring radius ($R$) is $A/R$ = 0.0975, 0.1025, and 0.200, respectively. In the lower right panel d) the green strips denote the spanwise vector when $A/R$ = 0.200, but it turns out to be rather insensitive to the amplitude of the Kelvin wave. 
}
\label{f.KWringUnitVectors}
\end{figure}

To determine the internal twist angle (up to a constant), and, therefore, also the spanwise vector, ${\bf N}$, we can either 
use Eq.~\eqref{e.itaSeifert} for the direction of the constant phase, or alternatively use Eq.~\eqref{e.itaVFM} for the classical case. 
Figure~\ref{f.intwistKWring} illustrates the internal twist angle $\theta$ when using the Seifert frame, Eq.~\eqref{e.itaSeifert},
for the case of $m=3$ at various Kelvin wave amplitudes. At the critical amplitude, $A_{\rm c}$, the angle
$\theta$ has a jump of $\pi$ at the azimuthal locations $\phi = \frac{\pi}{m}(2i-1)$, $i=1,\ldots,m$. 
This jump compensates the jump of $-\pi$ in the normal $\hat{\bf n}$ and bi-normal $\hat{\bf b}$ 
around the tangent, thus ensuring that the ${\bf N}$ vector varies smoothly as we move along the length of the vortex. 
The behaviour of the unit vectors $\hat{\bf n}$ and ${\bf N}$ is illustrated in Fig.~\ref{f.KWringUnitVectors} for few different 
Kelvin wave amplitudes and with $m = 3$. We note that in this case, the normal and bi-normal vectors make three complete rotations 
around the tangent only for amplitudes above $A_{\rm c}$.

\subsection{Numerical examples for single vortices}

\begin{figure}[!t]
\centerline{\includegraphics[width=0.99\linewidth]{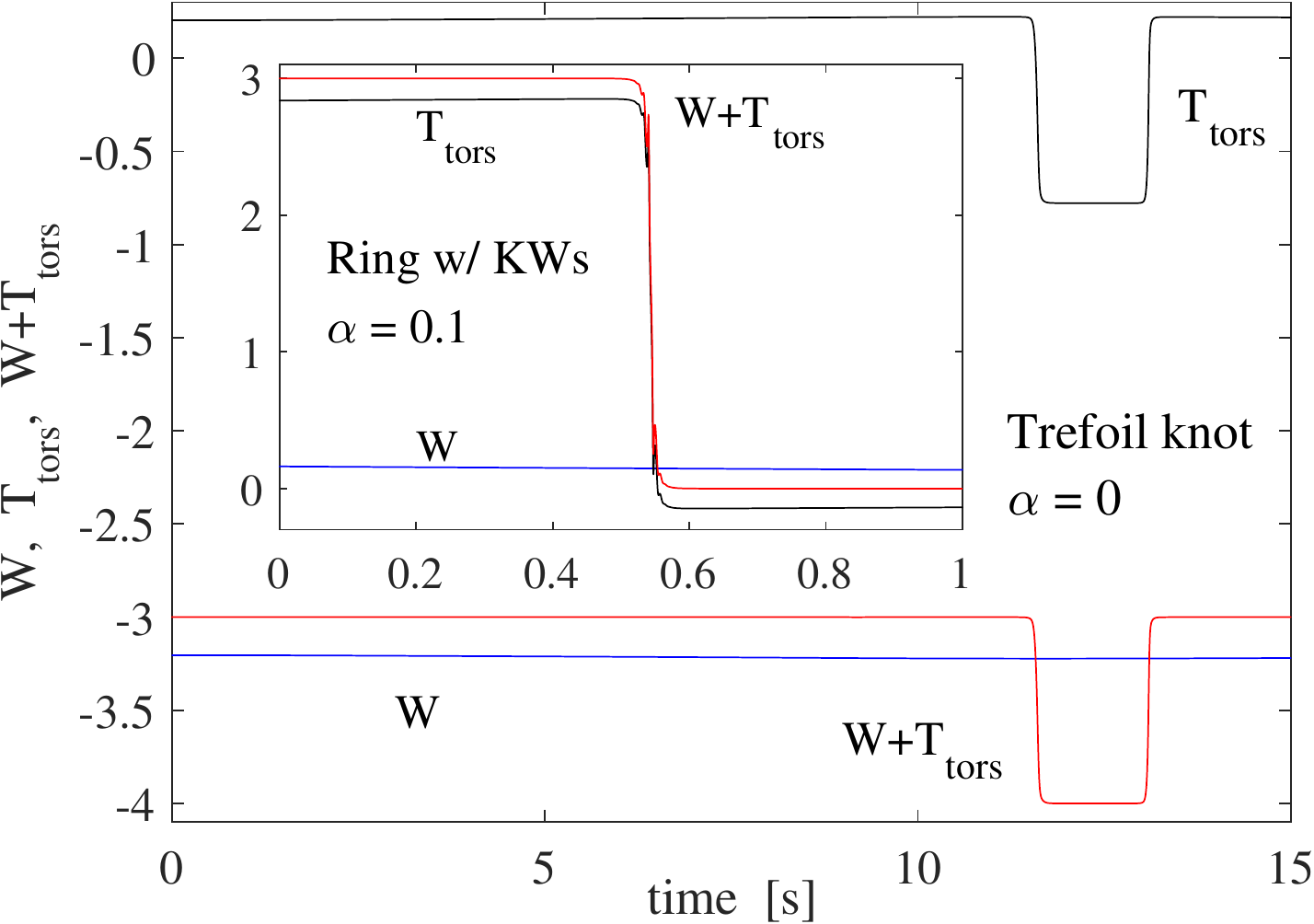}}
\caption{
Time dependence of writhe $W$ (blue) and torsion part of the twist $T_{\rm tors}$ (back), plus their sum (red). In the {\it main panel} the starting configuration is a trefoil knot. Here the temperature is zero (no dissipation) and the jumps in the torsion indicate that the internal twist $T_{\rm tw}$ has compensating jumps of $\pm$1 at times 11.6~s and 13.1~s, respectively. In the {\it inset} the initial configuration is a helical vortex ring of radius $R$ = 1 mm, with a Kelvin wave with $m = 3$ and amplitude $A/R$ = 0.12. The sudden decay of the torsion part of the twist $T_{\rm tors}$ is caused by mutual friction ($\alpha$ = 0.1, $\alpha'$ = 0) which slowly damps the Kelvin waves and drives the configuration through the (three simultaneous) inflection points.
}
\label{f.writhetorsKWringAndTrefoil}
\end{figure}

We will now focus on how the different contributions to helicity vary in time under a dynamical evolution of a vortex filament. 
In particular, we will be interested in an example where the vortex passes through an inflection point as it evolves. One example 
where inflection points appear is a breather solution\cite{Salman2013}. However, in that case the inflection points appear in 
pairs such that the torsion term $T_{\rm tors}$ is not effected and subsequently no changes occur in $T_{\rm tw}$. 
If mutual friction is added into the equations of motion (see Methods), the amplitude of the Kelvin waves decays with time. 
This is also accompanied by a decay of internal twist that can be clearly seen in our simulations 
An example of a simulation where mutual friction is included for an initial helical vortex ring is presented in the inset of 
Fig.~\ref{f.writhetorsKWringAndTrefoil}. The results illustrate how the torsion part of the twist suddenly drops by $3$ when 
the configuration passes through three inflection points. At the same time the internal twist that has an initial value of $-3$ 
jumps to zero such that the helicity remains zero. 

We have also modelled the dynamics of a trefoil knot at zero temperature. The main panel of Fig.~\ref{f.writhetorsKWringAndTrefoil} 
illustrates the behaviour of the writhe and torsion, plus their sum, in case of the trefoil 
knot where the initial configuration ($x$, $y$, and $z$ in mm's), parametrized by $t \in [0,2\pi)$, is given by
\begin{eqnarray} 
x &=& 0.41\cos{t}-0.18\sin{t}-0.83\cos{2t}-0.83\sin{2t} \nonumber \\ &-&0.11\cos{3t}+0.27\sin{3t} \,  \nonumber \\ 
y &=& 0.36\cos{t}+0.27\sin{t}-1.13\cos{2t}+0.30\sin{2t} \nonumber \\ &+&0.11\cos{3t}-0.27\sin{3t} \,  \nonumber \\
z &=& 0.45\sin{t}-0.30\cos{2t}+1.13\sin{2t} \nonumber \\ &+&0.11\cos{3t}+0.27\sin{3t}\, .
\end{eqnarray}

As for the vortex ring, we observe integer jumps in the torsion even though we are at zero temperature
due to the creation of inflection points as can be seen in Fig.~\ref{f.writhetorsKWringAndTrefoil}. These results 
demonstrate that the jumps of $\pm$1 in the torsion must be compensated by jumps of $\mp$1 in the internal twist in 
order to conserve the helicity. To ensure numerical accuracy, we have checked that the energy and momentum are conserved 
to within 0.1 percent throughout these simulations, up to the time of reconnections, which occur at times $>$ 60~s. 

\begin{figure}[!t]
\centerline{\includegraphics[width=0.8\linewidth]{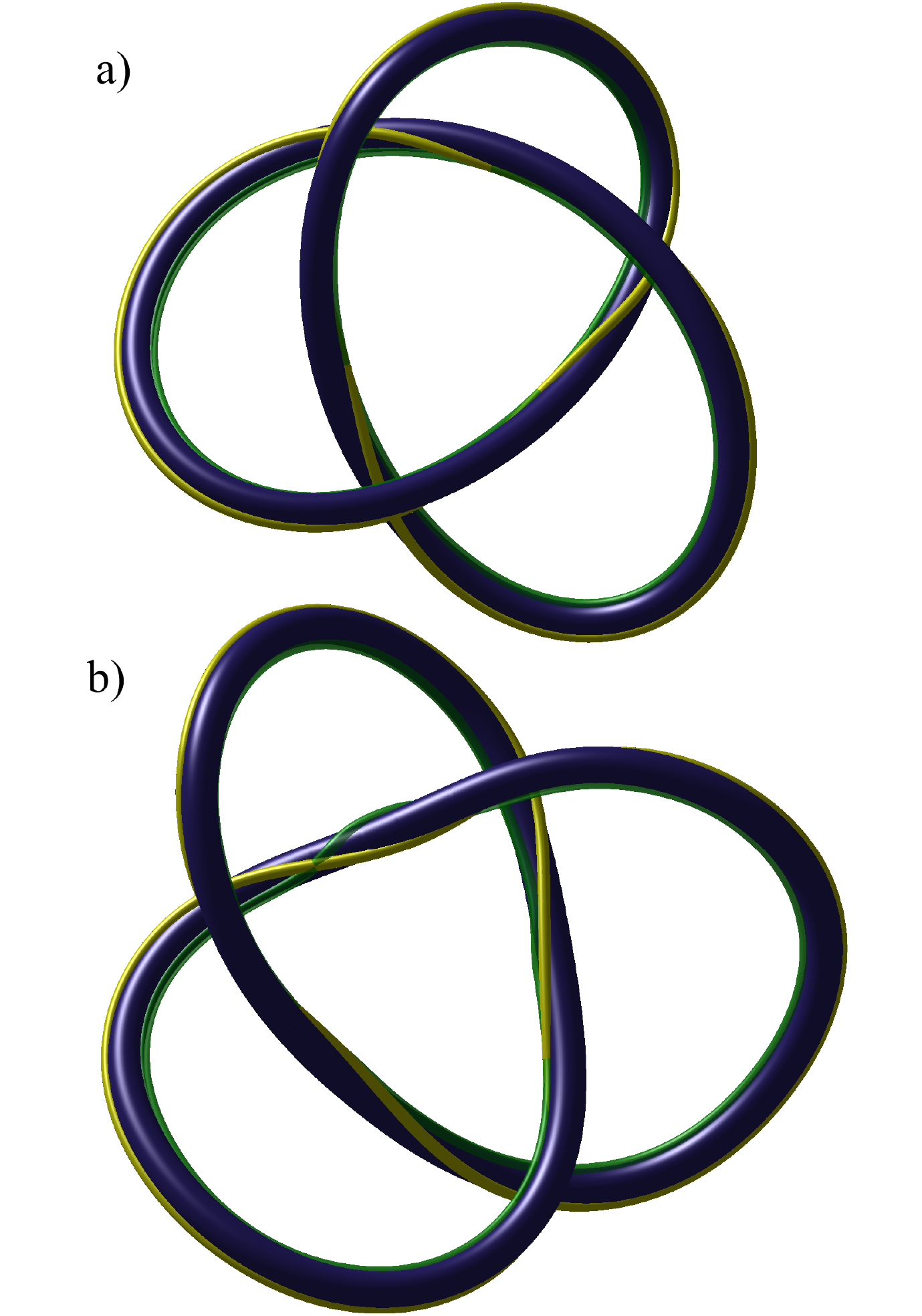}}
\caption{
Time evolution of a trefoil knot. Initial configuration is shown at the top panel a) where the internal twist $T_{\rm tw} = 3$. The bottom panel b) shows the configuration at 12.4~s where the internal twist is 4. The yellow stripes describe the rotation of the spanwise vector around the vortex when determined using the direction of constant phase (i.e. using Seifert framing). The green stripes describe the direction of the normal vector, which in the bottom figure makes an additional rotation around the vortex tangent, when compared with the upper figure.
}
\label{f.trefoilknot}
\end{figure}

The behaviour of the spanwise vector ${\bf N}$ in the Seifert frame and the normal vector $\hat{\bf n}$ during the trefoil 
knot dynamics is illustrated in Fig.~\ref{f.trefoilknot}. It is clear that following the formation of an inflection point 
(see Fig.~\ref{f.trefoilknot}b) that the Seifert frame, depicted by the yellow stripe, and the Frenet-Serret basis (green stripe) 
rotate in opposite directions. Therefore, whereas one varies covariantly, the other contravariantly which follows from 
their definitions. It is this property that results in the observed jump in the torsion at inflection points. 

\subsection{Vortex bundle}

The above results reveal the consequences of working in the Seifert framing. However,
a central question that remains is how would a quasiclassical limit of helicity emerge if we accept that helicity is trivially zero for a single superfluid vortex filament. From our considerations presented in Section \ref{s.hvfm}, we note that a classical vortex tube is well approximated by a bundle of vortex filaments that are aligned to the local vorticity field within the core of a classical vortex. In fact, such a quasiclassical limit of vorticity in superfluids has already been invoked to explain the measured Kolmogorov spectrum in superfluid turbulence\cite{Kozik2009}. 

\begin{figure}[!t]
\centerline{\includegraphics[width=0.99\linewidth,trim={0cm 4.2cm 0cm 4.3cm}]{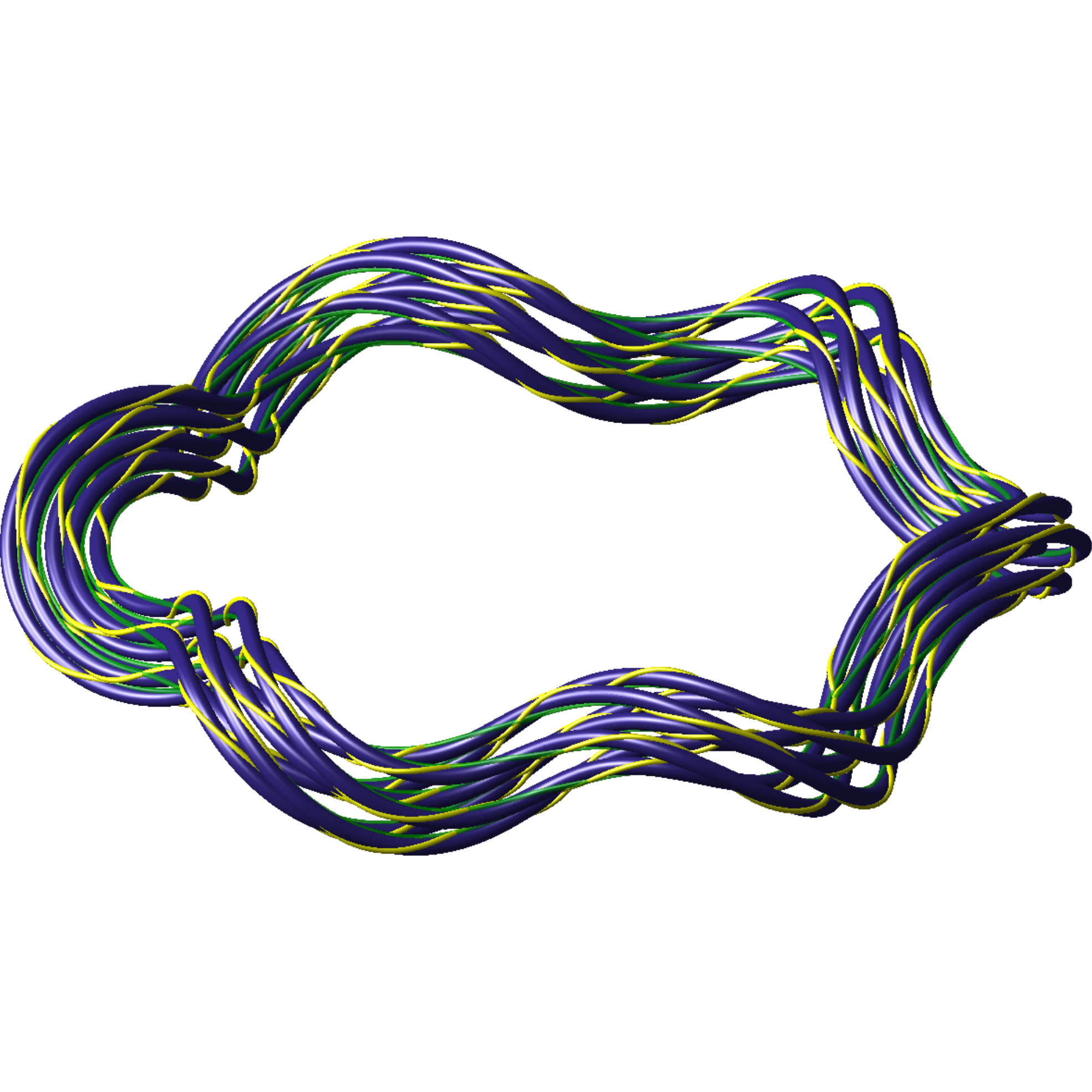}}
\caption{
Initial configuration for a twisted vortex bundle in a form of a vortex ring with Kelvin waves. The centremost vortex is a ring of radius $R$ = 1 mm with a Kelvin mode of $m$ = 6 and amplitude of $A = 0.1R$. The six outermost vortices are at a distance of 0.1 mm from the centre vortex and make 3 full rotations around (or linkings with) the centre one. The green stripes denote the direction of the normal unit vector, while the yellow stripes additionally indicate the direction of the constant phase around each vortex. Here the value of the constant phase is generally different on each vortex.
}
\label{f.bundleInitConf}
\end{figure}

\begin{figure}[!b] 
\centerline{\includegraphics[width=0.99\linewidth]{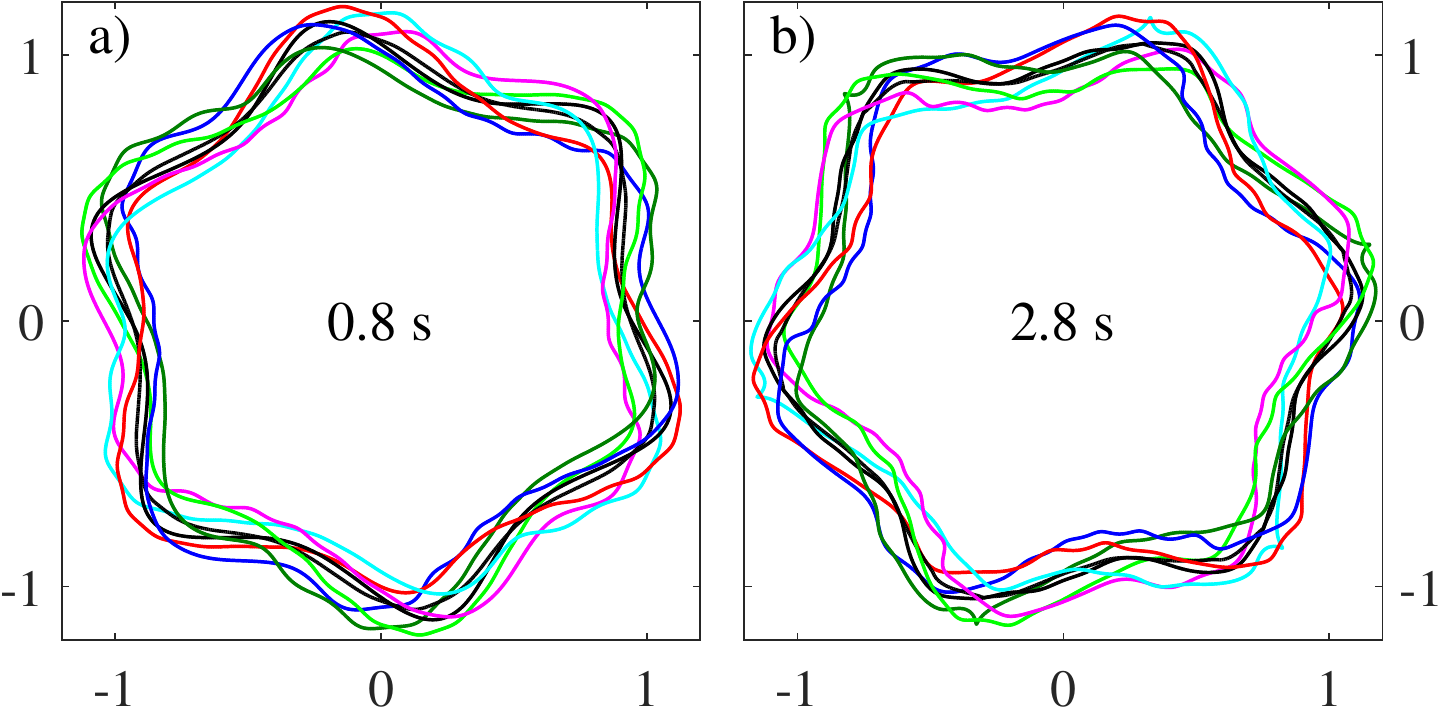}}
\caption{
Time development of the vortex bundle whose initial configuration is shown in Fig.~\ref{f.bundleInitConf}. The left panel a) describes the configuration at 0.8~s while the right panel b) is for 2.8~s. The different colors illustrate the different vortices such that the black solid line describes the vortex initially at the centre of the bundle. The dashed line describes the calculated centreline.
}
\label{f.bundleConfs}
\end{figure}

The bundles (or polarized discrete vortices) are created by stirring the superfluid. Remarkably, once formed, these polarized vortex bundles behave as coherent structures and can persist as has been illustrated from numerical simulations in Ref.~[\onlinecite{Alamri}]. Accepting that such coherent vortex bundles give rise to a quasiclassical vorticity field, it follows that the classical analogue of vorticity is defined by: 1) how the vortex lines within a vortex bundle twist around the centreline of the bundle; 2) how the centreline of the bundle writhes. So the quasiclassical notion of helicity is not tied to the Seifert frame  but emerges from the linking of vortex lines within the bundle. This is consistent with our observation in Section \ref{s.hvfm} that for a classical vortex tube, the twist is determined by the vorticity within the core and is unrelated to the Seifert frame. This implies that aside from the microscopic attributes of helicity for a single vortex filament, the quasiclassical limit should give rise to an emergent helicity conservation when the bundle evolves in the absence of reconnections.

\begin{figure}[!t]
\centerline{\includegraphics[width=0.99\linewidth]{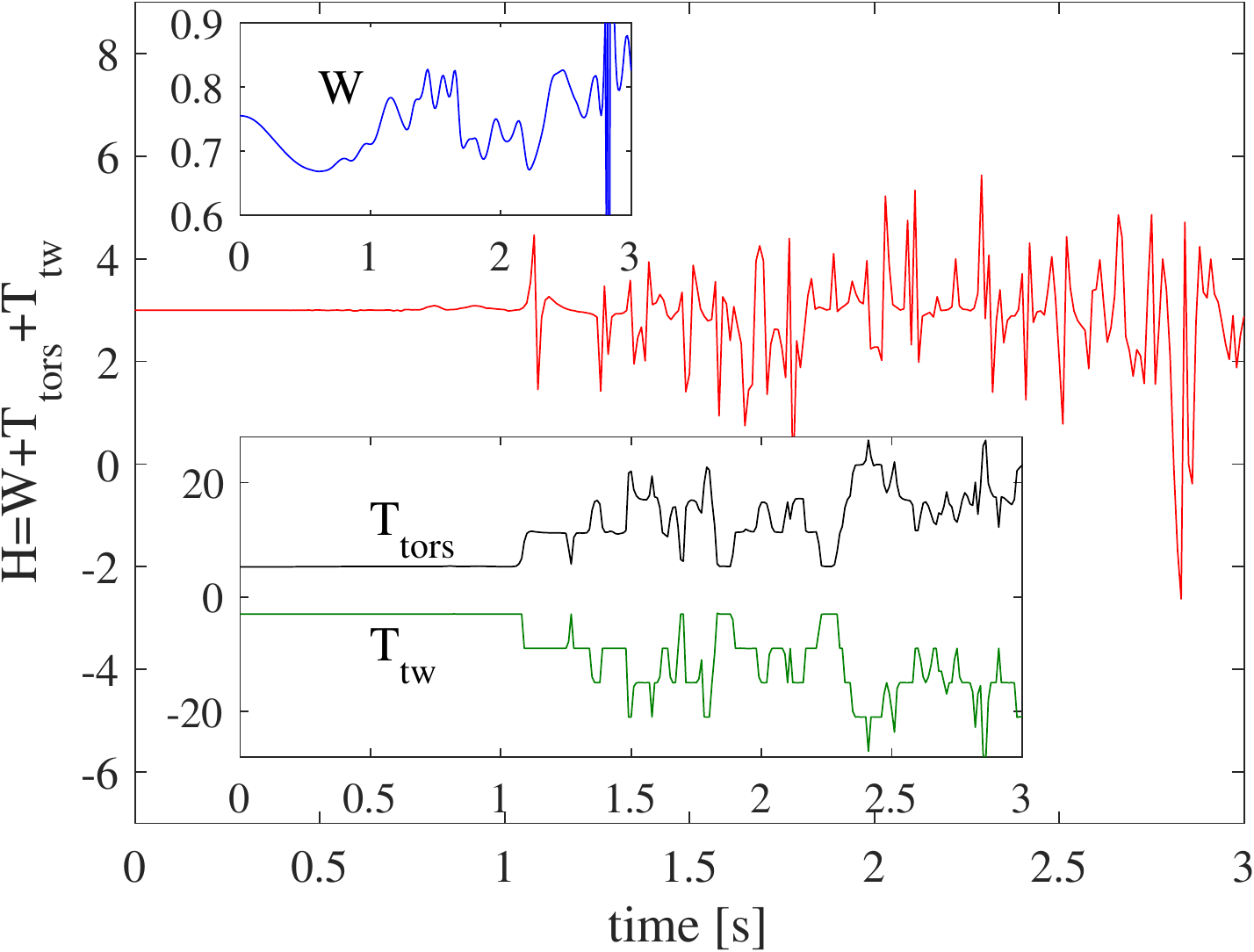}}
\caption{
Helicity and its components for the vortex bundle as a function of time. In the {\it main figure} the total helicity is plotted, while the {\it upper inset} illustrates the behaviour of the writhe and the {\it bottom inset} describes the time development of the torsion and internal part of the twist helicity.
}
\label{f.bundleH}
\end{figure}

To illustrate this behaviour, we have performed a numerical simulation of a bundle consisting of seven vortex rings perturbed with Kelvin waves and twisted 3 times around the bundle centreline, as shown in Figs.~\ref{f.bundleInitConf} and \ref{f.bundleConfs}. Details for of how the intial configuration was set up can be found in Sec.\ \ref{s.bundle}.
Upon integrating this configuration with a vortex filament model, we evaluated how the centreline writhe and the twist varied with time as shown in Fig.~\ref{f.bundleH}. As can be seen, although the writhe and twist vary, their sum remains essentially constant. The jumps in the centreline torsion are again due to inflections points in the centreline and are compensated by the internal twist. The noise in the total helicity is partly caused by the finite resolution which softens the jumps in the torsion but also because the bundle is loosing its coherence at later times. For example, in Fig.~\ref{f.bundleConfs}a, the different vortices remain as single-valued functions of the azimuthal angle. However, in Fig.~\ref{f.bundleConfs}b, steepening of the Kelvin waves on individual vortices destroys the coherent alignment of the vortices. Despite this, we note that since the helicity is approximately conserved and that the twist was evaluated from the rotation of the vortices within the bundle about the centreline, as described in Sec.\ \ref{s.bundle}, our results provide direct evidence of the emergence of a nontrivial quasiclassical helicity as an invariant of coherent vortex bundles.

\section{Conclusions}

When decomposing helicity into its constitent contributions, linking number, writhe, torsion, and internal twist, we find that the first three are completely prescribed by the instantaneous configuration of a vortex filament, whereas internal twist requires a spanwise vector to be defined. The sum of the first three contributions is an integer value and varies only when inflection points appear. In that regard, the torsion contribution to the helicity should always be retained when evaluating the helicity of a superfluid vortex which would otherwise never remain conserved under deformations of the vortex (even those that do not form inflection points) as was found in Ref.~[\onlinecite{Scheeler2014}] where only the writhe contribution was retained. 

At zero temperature the LIA (Localized Induction Approximation), where only the local term of Eq.~\eqref{e.BS2} is retained, possesses an infinite number of invariants, one of them being the integral 
of the torsion over the vortex length. This conservation of the torsion requires that inflection points 
appear in pairs, one contributing a change of $+1$ to $T_{\rm tors}$ and the other giving a change 
of $-1$. In other words the internal twist remains constant during the LIA dynamics. 

\begin{figure}[!t]
\centering
\includegraphics[width=0.7\linewidth]{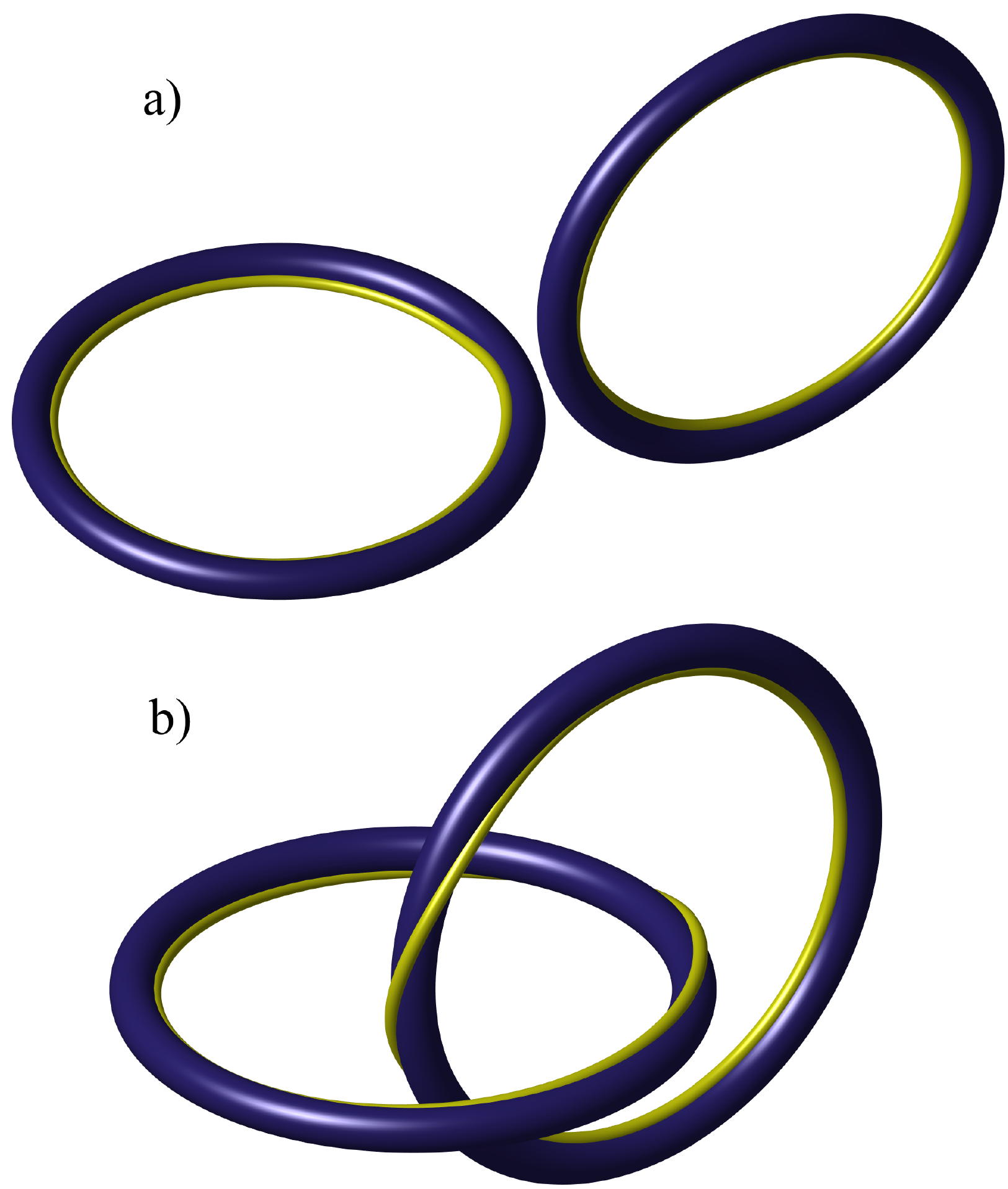}
\caption{
Rotation of the spanwise vector illustrated by the twist of the yellow stripe around the vortex. The internal twist angle is calculated using the Seifert frame, Eq.~\eqref{e.itaSeifert}, where the helicity zero. For unlinked rings a) the total twist is zero, while for linked case b) the spanwise vector makes one full rotation around each vortex, causing a nonzero twist. 
}
\label{f.spanwiserings}
\end{figure}

Inclusion of the internal twist restores helicity conservation even when inflection points are formed.
Defining the twist using the direction of the phase precludes the possibility of having an isolated twisted vortex ring. 
The phase of a twisted circular superfluid vortex would have a singularity at the centre. This is in contrast to the 
conventional definition of the twist. A circular vortex tube may be twisted, 
as illustrated by our bundle calculations where the tube is modeled by using seven vortices. The twisting 
of vorticity lines inside the tube induces a velocity along the vortex, but does not change the velocity 
potential outside the vortex tube. 

Whether the Seifert framing for superfluid vortices could be justified from first principles is left as an open question in this work. However, it appears to be a reasonable choice from at least three physical points. Firstly, it provides an unambiguous means to define the spanwise vector ${\bf N}$ that remains well-defined throughout the dynamics and in doing so it restores many of the characteristics of helicity that are known to hold for vortices in classical fluids. Secondly, it ensures helicity is conserved throughout the dynamics for superfluids even during reconnection events even though the different contributions would change to reflect the changing topology of the vortex lines.
A consequence of the Seifert framing is that this choice connects the linking with the internal twist. For example, two vortex rings which separately have zero twist, become twisted if linked (e.g. a Hopf link). For an illustration, see Fig.~\ref{f.spanwiserings}. Thirdly and most significantly, we have shown how a nontrivial quasiclassical limit of helicity can emerge that is independent of the Seifert framing.

\section{Methods}

\subsection{Dynamics with the filament model}\label{s.dvfm}

The dynamics of quantized vortices is typically modelled with the filament model\cite{Schwarz1985}
such that the vortices are considered as thin tubes with cylindrically symmetric cores where the circulation generates a classical Magnus force. Given the extremely small vortex core, we may ignore the small effective mass of the vortex although in general this effect can become more important at very low temperatures and on scales comparable to the core size. At finite temperatures, one additionally
needs to consider dissipation due to the mutual friction that couples the superfluid and normal components together. Setting the sum of the Magnus and mutual friction forces equal to zero results in the following equation of motion for the vortex lines:  
\begin{equation}
\frac{d{\bf s}}{dt} = {\bf v}_{\rm L} = {\bf v}_{\rm s}
+\alpha\hat{\bf s}'\times({\bf v}_{\rm n}-{\bf v}_{\rm s})
-\alpha'\hat{\bf s}'\times[\hat{\bf s}'\times({\bf v}_{\rm n}-{\bf v}_{\rm s})] \, .
\end{equation}
Here ${\bf v}_{\rm n}$ corresponds to the normal fluid velocity while ${\bf v}_{\rm s}$ corresponds to the superfluid velocity which is given by ${\bf v}_{\rm s} = {\bf v}_{\rm BS} + {\bf v}_{\rm Ex}$ where ${\bf v}_{\rm Ex}$ corresponds to some externally imposed irrotational superflow which we set equal to zero. Both the normal $({\bf v}_{\rm n})$ and superfluid $({\bf v}_{\rm s})$ velocities are evaluated at the vortex (centreline), ${\bf s}$, whereas derivatives are evaluated with respect to the arc length, such that $\hat{\bf s}'=\hat{\bf t}$ corresponds to the unit tangent vector. The two parameters $\alpha$ and $\alpha'$ represent mutual friction coefficients that depend on temperature and pressure. The effect of the corresponding terms is to damp out excitations such as Kelvin waves on the vortices. For calculations at finite temperature we also ignore the coupling from superfluid vortices on the normal component and take ${\bf v}_{\rm n} = 0$. This approximation is especially
suitable in superfluid $^3$He where the viscosity of the normal component is high.

Since we need to evaluate the superfluid velocity along the vortex centreline, we must introduce a cutoff to avoid the singularities in the superfluid velocity field arising from the divergent terms appearing in Eq.~\eqref{e.BS} for ${\bf v}_{\rm BS}$. This is accomplished by introducing a finite core size cut-off. This way the singularity in the azimuthal component of the velocity (as mentioned above, the tangential component of ${\bf v}_{\rm BS}$ is regular) can be avoided and the Biot-Savart velocity at a vortex point becomes
\begin{equation}\label{e.BS2}
{\bf v}_{\rm BS}({\bf s}) \!=\!
\frac{\kappa}{4\pi}\hat{\bf s}'\!\times\! {\bf s}'' \ln\left(\frac{2\sqrt{l_{+}l_{-}}}{e^{\beta}a_0}\right) \!+\! 
\frac{\kappa}{4\pi}\int^{'}\!\frac{({\bf s}_1-{\bf s})\!\times\! d{\bf s}_1}
{\vert {\bf s}_1-{\bf s}\vert^3}\, . 
\end{equation}
Here the derivates are again with respect to arc length $\xi$, and the integral omits the short segments 
$l_\pm$ around the point ${\bf s}$. The first term is called the local term and it produces a velocity along the bi-normal with amplitude proportional to local curvature $c = |{\bf s}''|$. The coefficient $a_0$ is the vortex core size and the $\beta$-parameter is related to the core structure, being 1/4 for a solid rotating core model (Rankine vortex) and 1/2 for a hollow core vortex. It is tuned such that the local term gives the same velocity for a vortex ring (or its local segment) as that of a classical vortex ring with a similar core structure. The approximation made for the local term is valid when $a_0 \ll l_\pm \ll 1/c$.

\subsection{Twist contribution for the vortex bundle}\label{s.bundle}

For the initial configuration we used a bundle of seven vortices where the centremost vortex, 
which initially is also the centreline, had a shape of the ring with a Kelvin-wave, Eq.\ \eqref{e.ringKW}. 
We set $R$ = 1 mm, and $A$ = 0.1 mm. The twist of the bundle was achieved by twisting the 6 outermost vortices 
by 3 times (corresponding to an angle of $6\pi$) with respect to the untwisted bundle, which can be obtained by setting the 6 outermost 
vortices at directions determined by the constant phase directions of $0, 2\pi/6, \ldots, 10\pi/6$, of the centre 
vortex (calculated when omitting the effect of the outer vortices). 
The twisting angle with respect to this untwisted configuration was chosen to depend linearly on
the arc-length along the centreline. The distance of the outermost vortices from the centre vortex
was 0.1 mm. Around 7000 points where used to discretise the vortices. These parameters were chosen 
primarily due to constraints set by the numerics.

During the time evolution of the vortex bundle the centreline was determined by averaging over the azimuthal locations of the seven different vortices. When deformations from a circle are not too large, this approximation is suitable. When evaluating the internal twist for the bundle, the direction of the different vortices with respect to the Serret-Frenet basis of the centreline was determined. In other words, the angle, $\theta$, appearing in Eq.\ \eqref{e.Neq}, was calculated for each vortex as a function of the length along the centreline by determining the crossing point of each vortex on the plane determined by $\hat{\bf n}$ and $\hat{\bf b}$.

As long as the bundle remains coherent the twist of the different vortices remains the same. The only exception was the centremost vortex, which initially determined the centreline. Because the centremost vortex may be very close to the centreline, the twist was determined using only the initially six outermost vortices.

\section*{Acknowledgements}
R.H. and N.H. acknowledge support from the Academy of Finland. 
H.S. acknowledges support for a Research Fellowship from the Leverhulme Trust under Grant R201540. We thank especially J. Karim\"aki, M. Krusius, R. Ricca, G.E. Volovik for fruitful discussions, comments and improvements to the paper. 
We also like to thank CSC - IT Center for Science, Ltd., for the allocation of computational resources. 




\begin{thebibliography}{99}
\bibitem{NoteIncomp}
We note that  in order to maintain full consistency throughout our formulation, it is important to work with the incompressible component of the phase (i.e.\ its divergence free projection). Therefore, whenever we refer to $\varphi$ in what follows, we will assume this represents only this component of the phase.
\bibitem{NoteSeifertFrame}
Not all Seifert frames are equivalent to surfaces of constant phase. However, here we only restrict 
to ourself to those which correspond to a constant phase and use the term Seifert frame.
\bibitem{noteContCore}
The requirement of constant twist velocity from the continuity equation is relevant to solid core model
(and similar ones) but naturally causes problems with hollow core model where the definition of the velocity 
inside the core becomes problematic.
\end{thebibliography}
\end{document}